\documentclass[prl,twocolumn,showpacs]{revtex4}
\usepackage{graphicx}
\usepackage{times}

\begin{document}
%\sloppy 
%\draft%revtex 3

    \title{Decoherence of Bose-Einstein condensates in microtraps}
%    \\[1ex]
    \author{C. Henkel$^1$ and S. A. Gardiner$^2$%
    }
%    \\[1ex]
%    \normalsize
    \affiliation{
%    \it
$^1$Institut f\"{u}r Physik, Universit\"{a}t Potsdam, 
    14469 Potsdam, Germany
    \\
%    Email: carsten.henkel@quantum.physik.uni-potsdam.de
%    \\
$^2$Clarendon Laboratory, Department of Physics, 
    University of Oxford, Oxford OX1 3PU, United Kingdom}
%   \\
    \date{16 September 2003}

%\maketitle%revtex 3

\begin{abstract}
%    {\bf Abstract.} %revtex 3
    We discuss the impact of thermally excited near fields on the
    coherent expansion of a condensate in a miniaturized 
    electromagnetic trap.
    Monte Carlo simulations are compared with a kinetic 
    two-component theory and indicate that atom interactions 
    can slow down decoherence. This is explained by a simple theory 
    in terms of the condensate dynamic structure factor.
\end{abstract}

\pacs{03.75.-b, 03.65.Yz, 05.40.-a}

\maketitle%revtex 4

The decoherence of atomic de Broglie waves is a key issue for applications
in atom interferometry and quantum information processing. 
It is particularly relevant for
integrated atom optics based on miniaturized hybrid electromagnetic surface
traps \cite{Folman02,Hinds01a}
because the atoms couple to a macroscopic, `hot' substrate nearby.
Loss processes due to spin flips driven by thermal magnetic near fields
have very recently been observed in the laboratory \cite{Jones03}, 
in agreement with predictions made by one of us \cite{Henkel99c}.
In this paper, we discuss a simple decoherence scenario for Bose-Einstein
condensed atomic matter waves in a quasi-one-dimensional microtrap.
This setup provides a realization of the standard model of 
environment-induced decoherence \cite{Zurek91} featuring two attractive
advantages: (i) the coupling to the environment can be microscopically
modelled in terms of the magnetic dipole interaction; (ii) due to atomic
interactions, the matter wave equation becomes nonlinear and novel features
are expected. We compare Monte Carlo simulations for the
condensate order parameter to a kinetic theory for the matter wave
coherence function and show that 
already for moderate interaction parameters,
a Bose-Einstein condensate is more robust with respect to a fluctuating
environment.

%\section{Model}

We consider an elongated trap similar 
to those formed above current carrying wires 
\cite{Folman02}. In the confinement dominated regime,
the matter waves can be described in a one-dimensional 
mean field approximation 
\cite{Fetter99} (units with $\hbar = m = 1$),
\begin{equation}
    {\rm i}\partial_{t} \psi = 
    - \frac12 \partial_{x}^{2} \psi + V(x,t) \psi
    + g |\psi(x,t)|^2 \psi,
\label{eq:GPe}
\end{equation}
where the interaction parameter $g = 2 \Omega_{\rm r} a / ( 1 -
1.46 a / a_{\rm r} )$ depends on the 
three-dimensional scattering length $a$, 
the radial confinement frequency $\Omega_{\rm r}$
and ground state size $a_{\rm r}$ \cite{Olshanii98}.
The density $|\psi(x,t)|^2$ is normalized to the total number 
of particles $N$. 
The potential $V(x, t)$ determines the dynamics in the axial 
direction. We assume that for $t < 0$, the atoms are confined in a 
harmonic trap with frequency $\Omega$ and occupy all
the zero-temperature condensate mode $\phi_{0}( x )$
\cite{noteInitFluct}. For $t\ge 0$, the axial 
confinement is switched off, 
the atoms expand, and we take into account their interaction with 
magnetic field fluctuations by letting $V(x,t)$ be a random potential.
Note that the radial confinement is kept constant.
Eq.~(\ref{eq:GPe}) thus describes the interplay between matter wave
interactions and time-dependent noise in an essentially
one-dimensional geometry. In contrast to previous work 
in the field of nonlinear random waves \cite{Konotop,Abdullaev}, our 
initial condition does not correspond to a self-contained soliton 
because we assume repulsive interactions $g > 0$.  Current 
experiments in wire traps have been hampered by the presence of a 
static field modulation that leads to the fragmentation of the 
expanding atom cloud \cite{Zimmermann02b,Lukin03}.  
This is neglected here and makes a direct comparison 
beyond the scope of our model.

%\subsection{Magnetic noise}

If we ignore spin flip processes for simplicity,
magnetic noise in a microtrap above a planar substrate translates
into a random potential with correlation function
\cite{Henkel99c,Henkel00b}
\begin{eqnarray}
    \langle V(x, t) V(x', t') \rangle &\approx&
    \gamma \, \delta(t - t') 
    C( |x - x'| )
    \label{eq:noise-correlation}
    \\
    C( |x - x'| ) &=& \frac{ l_{\rm corr}^2 }{
    l_{\rm corr}^2  + (x - x')^2}
,
    \label{eq:Lorentzian}
\end{eqnarray}
where $\gamma$ is the noise strength %is given by
% \(
%     \gamma = \mu^2 \mu_{0}^2 k_{\rm B}T_{\rm s} /(
%     16\pi \hbar^2
% \)
% \(    \varrho \,z_{\rm t} )
% ,
% \)
% involving magnetic moment $\mu$, substrate temperature
% $T_{\rm s}$ and resisitivity $\varrho$ and trap-surface
% distance $z_{\rm t}$. 
and the spatial correlation length $l_{\rm corr}$ 
is of the order of the microtrap height \cite{Henkel00b}.  
If the potential fluctuated only in time, 
$\gamma$ would correspond to the phase diffusion rate.
Typically, $1/\gamma$ is a few seconds in $\mu$m sized traps
\cite{Jones03,Henkel99c}.
%For a discussion of
%the height dependence $\propto 1/z_{\rm t}$ and corrections
%involving the substrate skin depth, see \cite{Henkel99c}.
%These corrections are required to describe quantitatively
%the spin flip losses observed in microtraps \cite{Cornell03a}.
In the frequency range relevant for our model
(up to $\Omega_{\rm r}$),
the noise spectrum is approximately flat
\cite{Henkel03a}.

%\subsection{Coherence functions}

\begin{figure}%[tbh]
%    \vspace*{30mm}
     \resizebox{55mm}{35mm}{%
     \includegraphics*{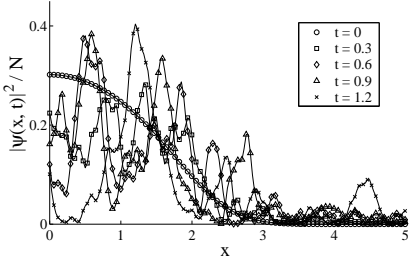}}%
%\vspace*{-03mm}
\caption[]{Expansion of a self-interacting Schr\"odinger field 
in a noisy potential,
single realization. The normalized spatial density is plotted at 
different times given in the inset.
%\\
Parameters in Eq.(\ref{eq:GPe}): interaction strength 
$gN = 10$, noise strength $\gamma = 1$, correlation length $l_{\rm 
corr} = 0.1$. Harmonic oscillator units with respect to the initial 
confinement frequency $\Omega$ are used:
$t \mapsto \Omega t$, $x \mapsto 
(\hbar/m\Omega)^{1/2} x$.
The numerical solution uses a discrete space-time grid with
time step $dt = 0.1$, and $2^{14}$ space points spaced 
$dx = 0.0294$ units.
% The time evolution is computed with a 
% split-operator algorithm.
}    
\label{fig:single-run}
\end{figure}

\begin{figure}%[tbh]
%    \vspace*{30mm}
     \resizebox{55mm}{35mm}{%
     \includegraphics*{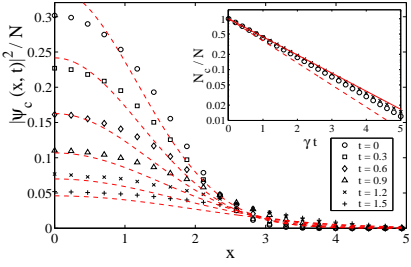}}%
%\vspace*{-03mm}
\caption[]{Normalized density profiles of the coherent
(noise-averaged) field for different expansion times.
%From top to bottom, the profiles are taken at $t = 0, \, 0.3, \ldots
%1.5$. 
Symbols: Monte Carlo results,
dashed lines: gaussian approximate solution to
Eq.~(\ref{eq:GPe-coherent}),
$|\psi_{\rm c}(x, t)|^2 = (N / (\sqrt{2\pi} u(t))
\, {\rm e}^{- \gamma t}
\exp[ - x^2 / (2 u^2(t) ]$ with $u(t)$ solving Eq.~(\ref{eq:force-u}).
%\\
Inset: coherent fraction (relative particle number). 
Crosses (open circles): Monte Carlo results for $\gamma = 1$
($\gamma = 0.1$). 
Dashed (solid) line: exponential decay
with decoherence rate $\gamma$ (renormalized rate $\gamma_{\rm eff}
= 0.82\,\gamma$).
Units and all other parameters 
as in Fig.~\ref{fig:single-run}.
}
\label{fig:becDecay}
\end{figure}

For a single, typical realization of the noise, the evolution
of the density $n(x, t ) = |\psi(x, t)|^2$ 
according to Eq.~(\ref{eq:GPe})
is shown in Fig.~\ref{fig:single-run}. A complicated fringe 
pattern appears due to the interference between the expanding 
condensate mode and the excitations generated by noise, with 
the fringe phase depending on the history of the noise. If
we average over the evolutions in an ensemble of noise potentials,
a smooth average field $\psi_{\rm c}( x, t ) \equiv
\langle \psi(x, t ) \rangle$
with a decaying weight emerges (Fig.~\ref{fig:becDecay}). 
This quantity would be revealed in a
(as yet hypothetical)
homodyne measurement of the Bose field, and
we shall call it the 
`coherent field' in the following.  
% Its absolute phase is 
% fixed by the phase of the initial wave function $\phi_{0}( x )$. 
Note the analogy to the condensate
order parameter in the symmetry breaking approach 
to Bose condensation when $N_{\rm c} \sim N$
\cite{Stringari99}.  
% Since the equation of
% motion for the atomic quantum field is essentially
% given by Eq.~(\ref{eq:GPe}), with classical fields replaced
% by operators, 
% classical simulations are also able to capture some aspects
% of quantum (and thermal)
% field fluctuations; for example, in the initial state one may
% include randomly chosen amplitudes for the lowest excitation modes of 
% the trapped condensate, see e.g.\ 
% \cite{Castin98,Davis01c}.

\begin{figure}%[tbh]
%    \vspace*{30mm}
     \resizebox{55mm}{35mm}{%
     \includegraphics*{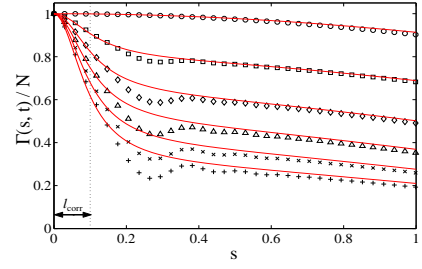}}%
%\vspace*{-03mm}
\caption[]{Spatially averaged coherence function [Eq.~(\ref{eq:def-cf})]
for different expansion times (values and symbol coding as in 
Fig.~\ref{fig:becDecay}).
%\\
Symbols: Monte Carlo results,
solid lines: kinetic theory [Eq.~(\ref{eq:result-cohfunc})] 
with $\Gamma_{\rm i}(s, 0) \equiv 0$, 
renormalized decoherence 
rate $\gamma_{\rm eff} = 0.82\,\gamma$ and noise correlation
length $l_{\rm eff} = 1.25 \, l_{\rm corr}$.
Other parameters as Fig.~\ref{fig:single-run}.
}
\label{fig:becCf}
\end{figure}

Another condensate definition, which is also applicable in 
$U(1)$ covariant theories, is based
on long-range order in the single-particle density matrix
\cite{Leggett01}. This 
quantity corresponds in our problem to the coherence function 
$\rho(x, x', t) \equiv \langle \psi^*( x, t) \psi( x', t) \rangle$.
Its decay with increasing separation $s = x' - x$ defines the
coherence length $l_{\rm coh}( t )$, a key concept
of decoherence theory \cite{Zurek91,Giulini}.
We introduce the spatial average
\begin{equation}
    \Gamma( s, t ) = \int\!{\rm d}x \, \rho(x, x+s, t)
,
\label{eq:def-cf}
\end{equation}
whose Fourier transform with respect to $s$ is the momentum 
distribution, averaged over many realizations.
This leads to $l_{\rm coh} \delta p
\sim 1$ where $\delta p$ is the width in momentum. 
The reduction of the coherence length (`decoherence')
is borne out in the results plotted in Fig.~\ref{fig:becCf}.
Long-range coherence is also visible: a fraction of 
the bosonic wave field is coherent across the full 
cloud size.
We shall see that this fraction can be identified with the coherent 
field $\langle \psi(x,t)\rangle$, reinforcing the 
analogy between the condensate order parameter and the
noise-averaged nonlinear Schr\"{o}dinger field.

%\section{Kinetic theory}

%\subsection{Basic equations}

In the two-component model of Bose-Einstein condensation,
the condensate evolves according to a nonlinear Schr\"{o}dinger
equation including loss terms and interactions with the
non-condensed density. The non-condensed component is described
by a suitable kinetic theory
\cite{Fetter99,Gardiner97c}.
We have adapted this model to our problem
by replacing the average with respect to the field's
density operator by the average over the evolutions in an ensemble
of random potentials. 
For a similar approach, see \cite{Kuklov00a}; the noninteracting case 
has been treated in \cite{Jayannavar82}. 
An essentially analytical solution has been 
obtained with two approximations: (i) we describe the 
fluctuations around the coherent field by the free space 
dispersion relation; (ii) we neglect in the nonlinear Schr\"{o}dinger equation
the interaction between the
coherent field and its fluctuations. The resulting equations are
\begin{equation}
    {\rm i}\partial_{t} \psi_{\rm c} = 
    - \frac12 \partial_{x}^{2} \psi_{\rm c} 
    + g |\psi_{\rm c}(x,t)|^2 \psi_{\rm c} 
    - \frac{ {\rm i}\gamma }{ 2 } \psi_{\rm c}
\label{eq:GPe-coherent}
\end{equation}
for the coherent field,
and the Boltzmann-type equation
\begin{eqnarray}
&&\left(
\partial_{t} + p \partial_{x}
\right) 
W_{\rm i}(x,p,t) =
\label{eq:transport}
\\
&&
\int\!{\rm d} p' \,
S_V(p - p')
   \left( 
   W_{\rm c}(x,p',t) + W_{\rm i}(x,p',t) - W_{\rm i}(x,p,t)
   \right)
\nonumber
\end{eqnarray}
for the Wigner representation $W_{\rm i}$ of the `incoherent' field. 
This is the Wigner transform of the fluctuating part of the
coherence function
\(
    \rho_{\rm i}(x,x',t) \equiv
    \langle \psi^*(x,t)\psi(x',t) \rangle
    -
    \psi^*_{\rm c}(x,t)\psi_{\rm c}(x',t)
.
%\label{eq:split-density-matrix}
\)
In Eq.~(\ref{eq:transport}), $W_{\rm c}$ is the Wigner transform
of $\psi_{c}^*(x,t)\psi_{c}(x',t)$, and the `collision integral'
involves the `cross section' \cite{Keller96}
\begin{equation}
    S_{V}( p - p' ) =
    \gamma \int\frac{ {\rm d}(x-x') }{ 2 \pi }
    C( x-x' ) \, {\rm e}^{ - {\rm i} (p - p') (x-x')  }
,
    \label{eq:xs}
\end{equation}
where $C( x - x' )$ is the normalized noise correlation 
function~(\ref{eq:Lorentzian}). 
When Eq.~(\ref{eq:transport}) is approximated by a Fokker-Planck
equation and $W_{\rm c} = 0$,
we essentially recover the decoherence model discussed by
W. H. Zurek \cite{Zurek91}.

%\subsection{Approximate solution}

The analytical solution to Eqs.~(\ref{eq:GPe-coherent}, 
\ref{eq:transport}) is derived using previous results 
for a noninteracting gas \cite{Jayannavar82,Henkel01a}.
The new ingredient is the nonzero average of the field
that enters the collision integral as a source term.
The basic idea of the analytical solution is to perform 
a Fourier transformation
with respect to both $x$ and $p$,
and to solve the resulting equation with the method of
% , reducing the collision 
% integral to a 
% product. The resulting equation is solved with the method of 
characteristics \cite{Arfken}. For the 
coherence function~(\ref{eq:def-cf}), this yields
\begin{eqnarray}
    \Gamma(s,t) & = &
\Gamma_{\rm c}(s,t)
+    
\Gamma_{\rm i}( s, 0 )\,
{\rm e}^{- \gamma t [1 - C(s)]} 
    \nonumber\\
    && 
    + \gamma C(s) \int\limits_{0}^{t}\!{\rm d}\tau \,
    {\rm e}^{- \gamma \tau [1 - C(s)]} \Gamma_{\rm c}(s,t-\tau)
.
    \label{eq:result-cohfunc}
\end{eqnarray}
The coherent field contribution,
$\Gamma_{\rm c}(s, t)$, can be found approximately using a 
time-dependent Thomas-Fermi profile \cite{Castin96} 
or a gaussian ansatz 
\cite{Lewenstein97a}. We follow the latter method
because it simplifies the Wigner transform and get
\begin{equation}
    \Gamma_{c}( s, t ) \approx
    N \,{\rm e}^{ - \gamma t }
    \exp\left\{ - \frac{ s^2 }{ 2 }
    \left( \frac{ 1 }{ 4 u^2(t) } + 
    \dot u^2(t)  \right) \right\}
    ,
    \label{eq:cond-cohfunc}
\end{equation}
where % $N$ is the total number of atoms and 
the spatial width $u(t)$ is the solution of
\begin{equation}
\ddot u  = - 
\frac{\partial}{\partial u} 
\left(
\frac{1}{8 \, u^2}
+ \frac{\tilde g( t )}{u}
\right)
.
\label{eq:force-u}
\end{equation}
The effective interaction strength is
$\tilde g( t ) = g N {\rm e}^{-\gamma t} / (4\sqrt{\pi})$,
and the initial condition minimizes 
\(
U_{\rm eff}( u ) = 1/(8 u^2)
+ \tilde g( t )/u + \frac12 \Omega^2 u^2
\)
where $\Omega$ is the initial trap frequency. 
% (In SI units, 
% \(
% U_{\rm eff}( u ) = \hbar^2/(8 m u^2)
% + \tilde g( t )/u + \frac12 m\Omega^2 u^2
% ,
% \)
% and similar changes apply to Eq.(\ref{eq:force-u}).
% In Eq.~(\ref{eq:cond-cohfunc}), multiply 
% $\dot u^2$ by the factor $(m/\hbar)^2$.)

The result~(\ref{eq:result-cohfunc}) of the kinetic theory splits 
into a contribution with long-range coherence (the first term
$\Gamma_{\rm c}$ extends across the entire cloud size) and an 
incoherent 
part (the second and third terms). For the initial conditions 
considered here, the second term vanishes so that the incoherent
fraction is not coherent beyond 
the correlation length of the noise potential, as intuitively
expected.
(Recall that the spatial correlation 
function $C(s)$ [Eq.~(\ref{eq:Lorentzian})]
decays on the scale $l_{\rm corr}$.)
% This confirms the 
% picture that above $l_{\rm corr}$,
% condensate excitations originate from sources with no fixed 
% phase relation required for interference.

%\subsection{Comparison to numerical simulations}

The simple kinetic theory outlined above captures qualitatively
the features observed in Monte Carlo simulations of Eq.~(\ref{eq:GPe}),
as shown by Figs.~\ref{fig:becDecay}, \ref{fig:becCf}.
We attribute deviations to the approximate treatment of interactions 
in the theory.

The density profile of the coherent field is not exactly 
gaussian (Fig.~\ref{fig:becDecay}) because for the chosen parameters, 
one already approaches the Thomas-Fermi regime. The coherent fraction
of particles $N_{\rm c}(t) / N = (1/N)\int\!{\rm d}x 
|\psi_{\rm c}(x,t)|^2$, however, shows an exponential decay as 
predicted by Eq.~(\ref{eq:GPe-coherent}). 
We find quantitative agreement with the kinetic theory 
when 
a `renormalized' scattering rate $\gamma_{\rm eff} < \gamma$ is 
used. 
Further simulation runs show that the ratio $\gamma_{\rm eff} / \gamma$ does 
not change significantly when the noise strength is reduced by an 
order of magnitude. 
A similar renormalization of the coherent field's loss rate
has been observed in simulations of nonlinear pulse
propagation with a randomly fluctuating phase mismatch
\cite{Lederer01a}.
An analytical approximation for $\gamma_{\rm eff} 
/ \gamma$ is derived below.

Similarly, the renormalization of the noise
correlation length to $l_{\rm eff} > l_{\rm corr}$ reproduces 
quantitatively both the short-range and long-range behavior of 
the coherence function (Fig.~\ref{fig:becCf}).  The 
oscillations originate from the Thomas-Fermi density kink 
at the condensate border that is not captured by  
our gaussian ansatz.

%\subsection{Renormalized decoherence rate}
%\label{s:gammaEff}

In a homogeneous condensate, atomic interactions suppress 
the generation of long-wavelength excitations by a spatially
modulated potential. This is described by the vanishing of the
dynamic structure factor $S[k; n_{\rm c}]$ in the limit $k \to 0$
% where $n_{\rm c}$ the condensate density 
\cite{Stringari99}. 
Let us consider for our problem the condensate as locally homogeneous
with a density $n_{\rm c}( x ) = |\psi_{\rm c}(x)|^2$. 
Including the structure factor in the scattering cross section, 
we suggest an improved approximation to 
the collision integral in Eq.~(\ref{eq:transport})
\begin{eqnarray}
    \partial_{t} W_{\rm i}(x, k, t) 
    \Big|_{\rm c \to i}
    &=&
\int\!{\rm d} k' \,
   S_{V}( k' )
   S[k'; n_{\rm c}( x )] 
   W_{\rm c}(x,k-k',t)
   \nonumber
   \\
   &\approx&
   S_{V}( k )
   S[k; n_{\rm c}( x )] 
   n_{\rm c}(x,t)
   ,
\label{eq:xs-with-sf}
\end{eqnarray}
where $S_{V}( k )$ [Eq.~(\ref{eq:xs})]
is the wavevector spectrum of the noise,
and the well-known Bogoliubov structure factor is 
\cite{Stringari99}
\(
    S[k; n_{\rm c}]
    =
    |k| / ( k^2 + 4 g n_{\rm c} )^{1/2}
.
\)
% which vanishes in the limit $|k|\xi \ll 1$ where
% $\xi = (4gn_{\rm c})^{-1/2}$ is the healing length.
In Eq.~(\ref{eq:xs-with-sf})
we have replaced the coherent field's momentum 
distribution by a $\delta$-function, assuming a width $u(t) 
\gg l_{\rm corr}$.  For the ideal gas, 
$S[k; n_{\rm c}( x )] \equiv 1$, and we recover 
Eq.~(\ref{eq:transport}). In the general case, we find
\begin{equation}
    \gamma_{\rm eff}( t )
    = 
    \frac{1}{N_{\rm c}(t)}
    \int\!{\rm d} x \,{\rm d}k\,
    S_{V}( k )
    S[k; n_{\rm c}( x )] 
    n_{\rm c}(x, t)
.
\label{eq:gammaEff-integral}
\end{equation}
The relevance of interactions 
is now determined by the competition between the 
width $1/l_{\rm corr}$ of $S_{V}( k )$ and the 
width $1/\xi(x)$ of the structure
factor, involving the local 
healing length $\xi(x) = (4gn_{\rm c}(x))^{-1/2}$.

\begin{figure}%[tbh]
%    \vspace*{30mm}
     \resizebox{55mm}{35mm}{%
     \includegraphics*{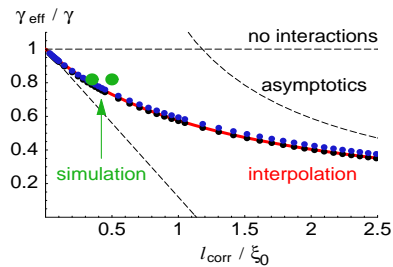}}%
%\vspace*{-03mm}
\caption[]{Renormalized decoherence rate 
$\gamma_{\rm eff}$
[Eq.~(\ref{eq:gammaEff-integral})]
for increasing interaction strength.
On the $x$-axis is plotted the ratio
$l_{\rm corr} / \xi_{0} \propto \sqrt{ g }$,
where $\xi_{0}$ is the healing length at the 
trap center, $\xi_{0} = 1/\sqrt{4gn_{\rm c}(x=0)}$.
Large dots: Monte Carlo data (see text).
Small lower (upper) dots: numerical
integration of Eq.~(\ref{eq:gammaEff-integral})
for a Thomas-Fermi (gaussian) density profile. 
Dashed lines: asymptotics for a Thomas-Fermi profile
at weak and strong interactions
[Eq.~(\ref{eq:gEff-asymptotics})],
thick solid line: interpolation~(\ref{eq:interpolation}).
The decoherence rate is normalized to its value
$\gamma$ in an ideal gas. 
}
\label{fig:gammaEff}
\end{figure}    

We have computed the integral~(\ref{eq:gammaEff-integral}) 
numerically for gaussian and Thomas-Fermi density profiles.
Very similar results are found (see Fig.~\ref{fig:gammaEff}) 
when the reduction factor $\gamma_{\rm eff} / \gamma$ is plotted
versus the ratio $l_{\rm corr} / \xi_{0}$ where the healing length 
$\xi_{0}$  is taken at the trap center. It turns
out that the time-dependence only appears via 
$l_{\rm corr} / \xi_{ 0 } \propto 1/\sqrt{ u(t) }$.
For the Thomas-Fermi profile, the $x$ integral can be evaluated 
analytically, leading to
\begin{equation}
    \frac{ \gamma_{\rm eff} }{ \gamma } =
    \frac{ 3 l_{\rm corr} }{4 }
    \int\limits_{0}^{\infty}\!{\rm d}k \,{\rm e}^{-k l_{\rm corr}}
    f( k \xi_{0} )
,
\label{eq:gammaEff-integral-2}
\end{equation}
where
\(
f(z) =
    z \left[ z + (1 - z^2)\arctan z \right]
    .
\)
Asymptotic analysis leads to the limiting behavior
(dashed lines in Fig.~\ref{fig:gammaEff})
\begin{equation}
    \frac{ \gamma_{\rm eff} }{ \gamma } =
\left\{
\begin{array}{ll}\displaystyle
1 - 0.88\,\frac{ l_{\rm corr} }{ \xi_{0} }
&
\mbox{for } l_{\rm corr} \ll \xi_{0}
,
\\[1ex]
\displaystyle
\frac{ 3\pi }{ 8 }
\frac{ \xi_{0} }{ l_{\rm corr} }
&
\mbox{for } \xi_{0} \ll l_{\rm corr}
.
\end{array}\right.
\label{eq:gEff-asymptotics}
\end{equation}
The ideal gas result $\gamma_{\rm eff} = \gamma$ is recovered
for $\xi_{0} = \infty$, but  
the next order correction already comes into play for a moderate
interaction parameter $g$, since $\xi_{0} \propto g^{-1/2}$. 
Very strong interactions, where $\xi_{0} \to 0$,
significantly slow down decoherence 
within our model. 
In the intermediate regime $\xi_{0} \sim l_{\rm corr}$, the 
asymptotics~(\ref{eq:gEff-asymptotics}) is quite inaccurate, 
but the interpolation
\begin{equation}
    \frac{ \gamma_{\rm eff} }{ \gamma } =
\frac{ A }{ A + 8 \,l_{\rm corr} / ( 3\pi \xi_{0})  },
\label{eq:interpolation}
\end{equation}
with $A \approx 1.15$ provides good agreement.
The decoherence rate extracted from the Monte Carlo 
results is also fairly well described by 
Eqs.~(\ref{eq:gammaEff-integral-2}, \ref{eq:interpolation}).
The two data points we have plotted correspond to two values for the 
healing length: based either on the numerically computed coherent
field density or on its Thomas-Fermi approximation.

Let us finally note that for large interactions or long expansion times,
the coupling to the incoherent fraction will no longer be negligible
compared to the random potential and our approximate solution will
lose accuracy. In this limit, we may expect that the noise-induced decoherence 
is replaced by an `open system dynamics' similar to the models discussed 
for a condensate at finite 
temperature \cite{Gardiner97c,Anglin97b}.

%\section{Conclusion and outlook}

To summarize, we have evaluated the impact of 
weak magnetic field fluctuations on coherent matter wave
dynamics in atom chips. 
A quantum kinetic theory for a degenerate trapped boson gas 
subject to noise has been worked out in the mean field
approximation and solved analytically,
neglecting the backaction of excitations onto the coherent
field. The comparison to numerical 
simulations demonstrates that interatomic interactions reduce the 
decoherence rate relative to the ideal gas. We have suggested
an explanation in terms of the structure factor of a
quasi-homogeneous system that leads to an accurate agreement
with the numerical data. Further investigations will address the
renormalization of the noise correlation length due to interactions
and the impact of finite temperature in the initial conditions.

\paragraph*{Acknowledgements.} ---
S. A. G. acknowledges support from the BEC 2000+ programme
of the European Science Foundation, the United Kingdom's 
Engineering and Physical Sciences Research Council 
and the Alexander von Humboldt foundation.
C. H. enjoyed travel support from the European Union
project ``Atom Chip Quantum Processor'' (contract no.
IST-2001-38863) and thanks V. V. Konotop, Arkadi Pikovski,
and the whole Quantum Optics group led by Martin Wilkens for 
encouraging comments.


\begin{thebibliography}{10}
\bibitem{noteInitFluct}
Thermal (and quantum) fluctuations of the atomic quantum
field can also be taken into account in the evolution according
to Eq.~(\ref{eq:GPe}) \cite{Castin98,Davis01c}; 
for example, by including randomly phased
amplitudes for the lowest condensate excitation modes 
in the initial conditions \cite{Dettmer01}.
\end{thebibliography}
\end{document}